\documentclass[10pt,letterpaper]{article}
\usepackage{opex3}
\usepackage{cite}
\usepackage{color}
\usepackage{graphicx}
\usepackage{amsmath}%
\usepackage{amsfonts}%
\usepackage{amssymb}

\begin{document}

\title{Photon pair generation in birefringent optical fibers}

\author{Brian J. Smith$^{1,2,*}$, P. Mahou$^2$, Offir Cohen$^2$, J. S. Lundeen$^3$, and I. A. Walmsley$^2$}

\address{$^1$Centre for Quantum Technologies, National University of Singapore\\
3 Science Drive 2, 117543 Singapore, Singapore\\
$^2$University of Oxford, Clarendon Laboratory, Parks Road\\
Oxford, OX1 3PU, United Kingdom \\ 
$^3$Institute for National Measurement Standards, National Research Council\\
1200 Montreal Road, Ottawa, Ontario, K1A 0R6, Canada}

\vskip-.3cm \parskip0pc\hskip2.25pc \footnotesize 
  \parbox{.8\textwidth}{\begin{center}\it*Corresponding author: \it\textcolor{blue}{\underline{b.smith1@physics.ox.ac.uk}} \rm \end{center} } \normalsize \vskip-.3cm



\begin{abstract} 
We study both experimentally and theoretically the generation of photon pairs by spontaneous four-wave mixing (SFWM) in standard birefringent optical fibers. The ability to produce a range of two-photon spectral states, from highly correlated (entangled) to completely factorable, by means of cross-polarized birefringent phase matching, is explored. A simple model is developed to predict the spectral state of the photon pair which shows how this can be adjusted by choosing the appropriate pump bandwidth, fiber length and birefringence. Spontaneous Raman scattering is modeled to determine the tradeoff between SFWM and background Raman noise, and the predicted results are shown to agree with experimental data.
\end{abstract}

\ocis{(270.0270) Quantum optics; (270.5585) Quantum information and processing; (060.4370) Nonlinear optics, fibers} 



\section{Introduction}

Optical quantum technologies such as quantum computation, quantum sensing, and quantum cryptography make use of the intrinsic quantum-mechanical behavior of non-classical light, such as quadrature-squeezed, precise photon-number, or entangled photon-pair states to surpass the performance afforded by classical light. Proposals have been made that use frequency entanglement between photon pairs to enhance timing measurements \cite{harris07, odonnell07, nasr08}, improve clock synchronization \cite{giovannetti01, giovannetti02, kuzucu05}, and cancel dispersion in interferometers \cite{franson92, steinberg92, Erdmann98, nasr03}. Frequency-time entangled photon pairs have also been used in fundamental tests of quantum mechanics itself, e.g. through violation of Bell inequalities \cite{franson89, shih93, kwiat93}. Advancement both in understanding of the character of quantum mechanics and facilitating development of new quantum technologies are underpinned by methods to prepare a variety of photonic quantum states, ranging from highly-entangled photon pairs to completely uncorrelated twin beams \cite{grice98, law00, grice01, valencia07, palmett07}. For these sources to be useful the complete quantum state of the output light must be controlled, including the spatial, spectral, and polarization degrees of freedom.

The most widespread sources of non-classical light have been based on nonlinear optical interactions, especially parametric downconversion (PDC) and four-wave mixing (FWM), both in the stimulated and spontaneous regimes. Historically, spontaneous PDC (SPDC) has been the work horse for quantum experiments due to the relative ease of experimental setup, inexpensive and widely available equipment, and a high yield of usable photon pairs. More recently, waveguide SPDC has been developed to take advantage of increased spectral flexibility, dramatically increased brightness and the ability to mode match into optical fibers \cite{Banaszek01, URen04, Gisin01}. Waveguided sources can be significantly brighter than bulk sources (per mW of pump power) due in part to the strong transverse spatial confinement of the pump (which increases the interaction strength) and longer guide lengths available for conversion (which increases the interaction length). However, SPDC in both bulk crystals and waveguides is limited by the natural dispersion of the nonlinear optical materials to spectral domains in which birefringent or quasi-phase matching can be accomplished \cite{uren05, nasr08}. Further, bulk crystal sources are usually multi-mode in character \cite{zhang07} and demand careful alignment and pump configuration \cite{valencia07, mosley08}. 

Recent progress in photon-pair generation by spontaneous FWM (SFWM) in optical fibers has gained much attention \cite{sharping01b, fan05, fan05b, rarity05, fulconis05, li05, chen05, nguyen06, fan07, fulconis07, li08, goldschmidt08, cohen09, mcmillan09, halder09}. Fiber sources are expected to be better suited for integration with optical-fiber and waveguide networks that are being developed for quantum information protocols \cite{clark09, politi08, matthews09, marshall09, smith09, politi09}, because of the similar spatial modes of the sources and networks. There have been three distinct approaches, aside from the scheme presented here, for the production of photon pairs using SFWM in optical fibers. These can be categorized by fiber structure and approach to phase matching: photonic-crystal fiber (PCF) \cite{sharping04, fan05, fan05b, rarity05, fulconis05, nguyen06, fan07, fulconis07, goldschmidt08, cohen09, halder09}, dispersion-shifted fiber (DSF) \cite{sharping01, fiorentino02, takesue05, chen05, li05, liang06, li08, dyer08, dyer09}, and standard single-mode fiber (SMF) \cite{altepeter08, hall09, hall09b}. Our approach uses birefringent single-mode fiber (BSMF). The majority of the three previous methods all share a common aspect in that they utilize waveguide dispersion in a symmetric guide to achieve phase matching. The critical parameter for this is the fiber zero-dispersion wavelength (ZDW) \cite{stolen82, agrawal06}. In principle, the ZDW can be controlled by manipulation of the fiber transverse-spatial structure in PCFs \cite{russell06} and addition of impurities in DSFs \cite{sharping01}. The latter can only tune the fiber ZDW to wavelengths longer than the innate fiber ZDW. SFWM in SMFs harnesses the natural ZDW of silica fibers near 1310\,nm using modulation instability (MI) to achieve phase matching. To date, the DSF and SMF approaches have produced photons in the near-infrared spectral domain, where high-efficiency single-photon detectors are not readily available. Furthermore, these techniques have been plagued by Raman scattering that adds noise to the signal of interest. Suppression of Raman noise can be accomplished by cooling the optical fiber \cite{takesue05, lee06, dyer08, dyer09, hall09b}, although this adds a layer of experimental difficulty. Thus, using the ZDW to achieve phase matching restricts the ability to tune the pump laser and wavelengths at which photons can be produced. In contrast, the scheme presented here is flexible in the pump wavelength allowing photon pairs to be generated at arbitrary wavelengths. Instead of waveguide dispersion, it makes use of birefringence in an asymmetric guide to achieve phase matching, as previously demonstrated in the large gain regime \cite{stolen81, murdoch95}. 

Besides the potential to tune fiber dispersion, PCFs have high transverse spatial mode confinement ($\approx 2.5 \mu m$ mode diameter) compared to standard optical fibers designed for visible light ($\approx 5 \mu m$ mode diameter), making the effective interaction stronger by roughly a factor of four. However, the exotic fiber structure that gives rise to the controlled dispersion in PCF leads to transverse spatial modes that are not well matched with integrated photonic circuits \cite{clark09, politi08, matthews09, marshall09, smith09, politi09} in which the produced photons are to be used. This limits the coupling efficiency between PCF and SMF, currently 65\%, that can be achieved \cite{fulconis05}. Transverse-spatial mode mismatch does not significantly affect DSF, SMF, and BSMF techniques. Furthermore, unavoidable fluctuations in PCF manufacturing process may have a large effect on their dispersion due to the small core size and strong dispersion dependence on the spatial structure of the fiber, leading to a random distribution of two-photon spectral states from the same fiber draw \cite{cohen09, halder09}. Indeed, the ability to create PCF with desired dispersion is still not a well-developed technology, and thus experimentalists must search the catalog of currently available fibers to find the fiber with dispersion that comes closest to their needs. Then the pump central wavelength and bandwidth must be carefully tuned to achieve the joint spectral state desired \cite{cohen09, halder09}. Dispersion of standard fibers is approximately equal to that of their constituent material in bulk when the wavelengths involved are smaller than the core diameter, and thus do not suffer as much from small fluctuations. 

Nonclassical light sources based on FWM in standard birefringent fiber represents a useful alternative to PCF, DSF, SMF and bulk PDC sources. Photon pairs generated in this way are better mode matched to integrated optical networks compared to both bulk and PCF sources. Birefringent fiber has nearly 30 years more development than PCF, which makes them significantly more cost effective, and more likely to be uniform from draw to draw. By utilizing fiber birefringence for phase matching, photons can be created far from the Raman scatter, but still in the visible and near infrared where currently available single-photon-counting detectors can be used. The added control over the SFWM spectral state allowed by fiber birefringence can be tuned through stress induced birefringence, e.g. by transverse compression.

In this paper theoretical and experimental results showing how SFWM in standard birefringent fibers can be used to produce a range of two-photon spectral states are presented. The high heralding efficiency, low Raman background, high brightness, ability to tune the joint spectral state, and readily available components make this source useful in current applications. We begin by laying out the theoretical description of SFWM in birefringent fibers. A discussion of Raman scattering and its role in SFWM follows. Finally, the results of experiments to characterize SFWM in off-the-shelf birefringent fibers are presented.


\section{Spontaneous four-wave mixing in birefringent fibers: Theory}
\begin{figure}[h]
\centering\includegraphics[width=0.85\textwidth]{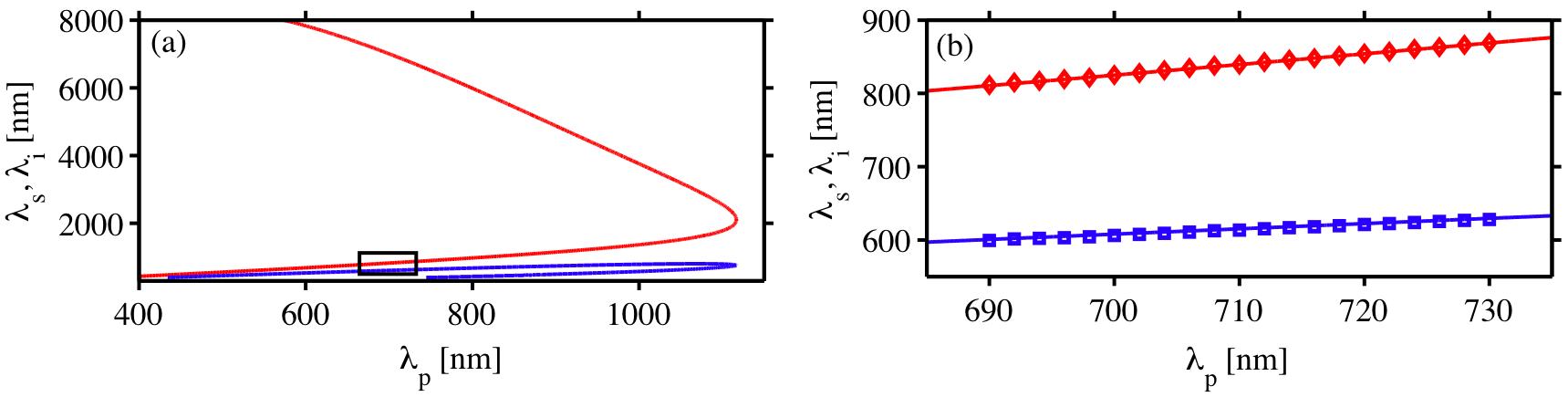}
\caption{Theoretical and experimental birefringent phase-matching contours as a function of the pump central wavelength, $\lambda_p$. (b) Magnified view of inset box in (a). The upper red line ($\diamond$) shows the idler ($\lambda_i$) theory (experiment) and the lower blue line ($\Box$) shows the signal ($\lambda_s$) theory (experiment) for the Fibercore HB800G fiber with $L =$ 0.2\,m and $\Delta n = 4.3 \times 10^{-4}$. Experimental pump power was 15\,mW, and theory plots neglect the $\gamma P$ contribution in Eq. (\ref{eq:bi-phasematching}). Experimental uncertainties are smaller than the data symbols.}
\label{fig:bi-phasematch}
\end{figure}

In the process of spontaneous four-wave mixing two photons from potentially two different pump pulses ($p1$ and $p2$) are converted into a pair of daughter photons, known as the signal ($s$) and idler ($i$), moderated by a third-order nonlinear optical material. This process occurs when energy is conserved and is most probable when the wave-vector mismatch is zero, i.e.
\begin{equation}
\omega_{p1} +\omega_{p2} = \omega_s + \omega_i,
\label{eq:e-cons}
\end{equation}
\begin{equation}
\Delta k = k_{p1}(\omega_{p1}) + k_{p2}(\omega_{p2}) - k_s(\omega_s) - k_i(\omega_i) + (1-B)\gamma (P_1 + P_2 + 2\sqrt{P_1P_2}) = 0,
\label {eq:phasematching}
\end{equation}
\noindent where $\omega_j$ and $k_j(\omega_j)$ ($j= p1,p2,s,i$) are the angular frequency and wave vector evaluated at frequency $\omega_j$ of mode $j$, $\gamma$ is the nonlinear parameter of the fiber, $B$ is related to the polarization of the pumps relative to the signal and idler fields, and $P_1(P_2)$ is the peak power of pulse pump 1 (2). Frequencies that satisfy both energy conservation, Eq. (\ref{eq:e-cons}), and zero wave-vector mismatch (i.e. $\Delta k = 0$), Eq. (\ref{eq:phasematching}), are said to be phase matched. There are several configurations in which phase matching can be achieved \cite{palmett07}. This paper focuses on the case in which a single pump pulse polarized along one of the principal axes of a birefringent fiber, either the fast (f) or slow (s) axis, generates signal and idler photons polarized along the opposite fiber axis, s or f, respectively, as previously examined in PCF \cite{cohen09, halder09} and in standard birefringent fibers in the strong pump regime \cite{stolen81, murdoch95}. In this cross-polarized birefringent phase-matching case, the wave-vector mismatch can be written to explicitly bring out the contribution of the fiber birefringence, $\Delta n$, as
\begin{equation}
\Delta k = 2 \frac{\omega_p}{c} n(\omega_p) - \frac{\omega_s}{c} n(\omega_s) - \frac{\omega_i}{c} n(\omega_i) + 2 \Delta n\frac{\omega_p}{c} + \frac{2}{3} \gamma P.
\label{eq:bi-phasematching}
\end{equation}
\noindent Here the dispersion of the two principal axes is modeled with the same material dispersion, given by the refractive index $n(\omega)$, and a constant offset, $\Delta n$, added to the slow axis, along which the pump is assumed to be polarized. The factor of $2/3$ arises from the self- and cross-phase modulation between the various modes ($B=2/3$), and $P$ is the pump peak power \cite{agrawal06}. 
Note that this approximation can be justified when far from the zero-dispersion wavelength, where the material dispersion dominates cross-polarized phase matching \cite{stolen81}, and when the birefringence $\Delta n$ changes little over the wavelength range of interest. The phase-matching contours, that is signal and idler frequency pairs as a function of the pump central frequency that satisfy energy conservation and phase matching ($\Delta k = 0$), are plotted in Fig. \ref{fig:bi-phasematch} for birefringent phase matching in silica fiber, Eq. (\ref{eq:bi-phasematching}). To make this graph, the Sellmeier equations for silica \cite{kasap06} are used, with $\Delta n = 4.3 \times 10^{-4}$, the pump polarized along the slow axis, and assuming self- and cross-phase modulation terms are negligible. Note that phase matching is not possible when the pump is polarized along the fast axis in the normal dispersion regime considered here \cite{stolen81, stolen82}.

\begin{figure}[h]
\centering\includegraphics[width=\textwidth]{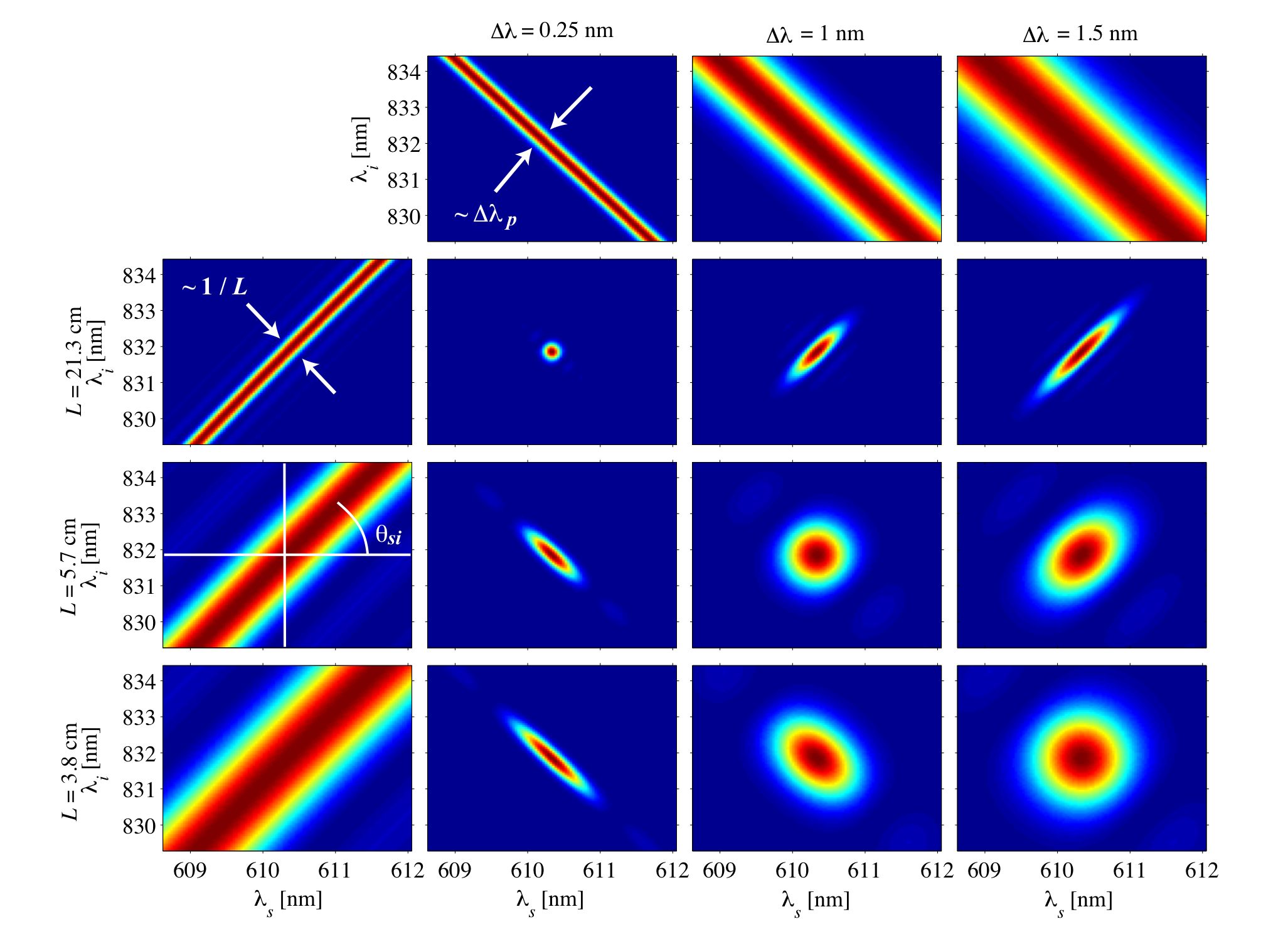}
\caption{Modulus squared of pump spectral amplitudes for different pump bandwidths (top row of plots, $\Delta\lambda =$ 0.25, 1, and 1.5\,nm FWHM), phase-matching function for different fiber lengths (left column of plots, $L = $ 21.3, 5.7 and 3.8\,cm) and theoretical joint spectra (lower-right 3$\times$3 array of plots) as a function of signal and idler wavelengths for birefringent phase matching. Plots are modeled using a pump central wavelength of 704\,nm, the bulk silica dispersion, $\Delta n = 4.3 \times 10^{-4}$, as measured for the Fibercore HB800G fiber, and neglecting the $\gamma P$ contribution in Eq. (\ref{eq:bi-phasematching}). The pump bandwidths and fiber lengths were chosen to display the variety of spectral states that can be realized using the birefringent phase-matching technique. }
\label{fig:joint-spec}
\end{figure}

The two-photon state of the signal and idler modes of the electromagnetic field emitted from the fiber, ignoring the vacuum component, can be written as \cite{palmett07}
\begin{equation}
|\psi \rangle = \int\int d\omega_{s}\,d\omega_{i}\,f(\omega_{s},\omega_{i})|\omega_{s}\rangle_s|\omega_{i}\rangle_i,
\label{eq:state}
\end{equation}
\noindent where $f(\omega_{s},\omega_{i})$ is the two-photon spectral amplitude of the signal and idler, and $|\omega_{s}\rangle_s|\omega_{i}\rangle_i$ is a two-photon state with frequencies $\omega_s$ and $\omega_i$. The transverse-spatial and polarization modes, assumed independent of frequency and factorable between each mode ($s$ and $i$), are omitted from Eq. (\ref{eq:state}) for clarity. The two-photon spectral amplitude can be written in terms of the pump spectral amplitude $\alpha( \omega)$ and the phase-matching function $\phi(\omega_{s},\omega_{i})$ \cite{palmett07}, as
\begin{equation}
f(\omega_{s},\omega_{i}) = \int d\omega' \alpha(\omega') \alpha(\omega_{s} + \omega_{i} - \omega') \phi(\omega_{s},\omega_{i}).
\label{eq:jsa}
\end{equation}
\noindent The phase-matching function, $\phi(\omega_s,\omega_i) = {\rm{sinc}}(\Delta kL/2){\rm{exp}}(i\Delta kL/2)$, depends on fiber length, $L$, birefringence, $\Delta n$, and dispersion as can be seen from the definition of the wave-vector mismatch, Eqs. (\ref{eq:phasematching}) and (\ref{eq:bi-phasematching}). The two-photon spectral state, $f(\omega_{s},\omega_{i})$, is completely determined by the pump-pulse spectral amplitude and fiber properties through the phase-matching function. For cross-polarized phase matching in silica fiber the birefringence, $\Delta n$, and pump central frequency, $\omega_{p0}$, determine the central frequencies of the signal and idler, positioning the joint-spectral state of the photon pair. The overall shape of the joint-spectral state is governed primarily (i.e. to first order in detuning from the ideal phase-matched frequencies) by the pump bandwidth, $\Delta \omega_p$, the fiber length, $L$, and the angle, $\theta_{si}$, the phase-matching function makes with the $\omega_s$ axis. The relation between pump amplitude, phase-matching function and joint spectrum is illustrated in Fig. \ref{fig:joint-spec}. Here the pump intensity, $|\alpha((\omega_s+\omega_i)/2)|^2$ (assuming a Gaussian pump), phase-matching probability, $|\phi(\omega_s,\omega_i)|^2$, and the joint-spectral probability, $|f(\omega_s,\omega_i)|^2$ are plotted for three different pump bandwidths and three different fiber lengths. These figures are calculated using Eqs. (\ref{eq:bi-phasematching}) and (\ref{eq:jsa}). Note that the width of the pump spectral amplitude is governed by the pump bandwidth, $\Delta \omega_p$, the width of the phase-matching function is proportional to the inverse fiber length, $1/L$, and the angle that the phase-matching function makes with the $\omega_s$ axis is $\theta_{si} = - \tan^{-1}(\tau_s / \tau_i)$. This angle is related to the group velocities of the signal, idler and pump modes, and thus to the higher-order dispersion properties of the fiber. Here $\tau_{s(i)}$ represents the group-velocity mismatch between the pump and signal (idler), given by
\begin{equation}
\tau_{s(i)} = L [ k_{p}'(\omega_{p0}) - k_{s(i)}'(\omega_{s(i)0}) ],
\label{gvmismatch}
\end{equation}
\noindent where $k_j'(\omega_{j0}) = {\rm{d}}k_{j} / {\rm{d}}\omega|_{\omega = \omega_{j0}}$ is the first-order derivative of mode $j$'s wave vector with respect to frequency evaluated at the central frequency of mode $j$.

Of particular interest for photonic quantum computation is the ability to herald pure-state single photons \cite{valencia07, mosley08, cohen09, halder09, soeller09}. For this the joint-spectral amplitude must be factorable, i.e. $f(\omega_s,\omega_i) = g(\omega_s) h(\omega_i)$ \cite{grice98, grice01, uren05, mosley08}. Approximating the phase-matching function by a Gaussian, ${\rm{sinc}}(\Delta k L/2) \approx \exp(-r (\Delta k L/2)^2)$, ($r = 0.193$), allows approximate determination of the regime in which a factorable joint-spectral state can be achieved for a given fiber. The pump bandwidth should be appropriately chosen to eliminate correlations, balancing pump and phase-matching bandwidths. In this Gaussian approximation a factorable joint-spectral amplitude can be attained for a pump bandwidth ($1/e$ spectral amplitude) given by 
\begin{equation}
\Delta \omega_p \approx \sqrt{2/(r|\tau_s\tau_i|)}.
\label{eq:pumpbandwidth}
\end{equation}
Note that this only works when the group-velocity mismatch terms ($\tau_j$) have opposite signs, that is, one mode must be faster and the other slower than the pump \cite{grice01, palmett07}, so that the phase-matching angle, $\theta_{si}$, lies in the range $[0,\pi/2]$. In silica BSMF, this angle is always less than $90^{\circ}$ in the visible and near infrared making it possible to create heralded single photons in pure quantum states over a broad spectral range.

In actual experiments, the phase-matching function is not a Gaussian, but a sinc function, with significant amplitude far from the central peak. This causes the joint-spectral state to contain more than a single set of signal and idler Schmidt modes \cite{mosley08, mosley08b}, implying a heralded photon from a realistic source to be produced in a mixed rather than pure state. To obtain a pure state in an experiment the tails of the sinc function can be spectrally filtered out leading to the theoretical purity limit of 1. This purity increase comes at the expense of overall production rate and heralding efficiency. Using our model of birefringent-phase-matched SFWM for $L = 10$\,cm and $\Delta n = 4.3 \times 10^{-4}$, with a Gaussian pump centered at 704\,nm, 0.5\,nm bandwidth (spectral intensity) full-width at half maximum (FWHM), the purity can be increased from approximately 0.86 to 0.99 using 90\% transmittance hard-edge filters centered on the signal and idler wavelengths with 0.7\,nm and 1.4\,nm spectral bandwidths FWHM, respectively. The production rate is reduced by 18\% and heralding efficiencies are reduced by about 10\% in each beam.

In addition to the complication arising from the wings in the phase-matching function, there is a another subtlety that arises in realistic experiments -- the pump spectral shape may not necessarily be well approximated by a Gaussian in a given experiment. For example, in Ref. \cite{cohen09}, the pump spectrum took on a nearly rect-function profile due to the nature of the spectral filtering used to prepare the pump. Such a spectral amplitude (with constant phase) leads to significantly reduced purity for heralded photons compared to a Gaussian profile, by roughly 16\%, due to the asymmetry introduced by the pump ``hard" edges. It is therefore crucial to take care when modeling and performing actual experiments to create the desired two-photon state.


\section{Raman scatter in SFWM}

Another nonlinear process, Raman scattering, can occur simultaneously with SFWM and produce photons at the signal and idler wavelengths. The process of Raman scattering can be viewed as a pseudo-three-wave mixing process, in which one of the waves is a vibrational excitation of the material, an optical phonon. There are two ways in which this can occur (up or down conversion). In the quantum picture the two Raman scattering processes can be described as the splitting of a pump photon with frequency $\omega_p$, into a Stokes photon of frequency $\omega_S = \omega_p-\Omega$, and a phonon of frequency $\Omega$, or the annihilation of a pump photon and a phonon of frequency $\Omega$, creating an anti-Stokes photon with frequency $\omega_A = \omega_p+\Omega$. At room temperature and with small detuning ($\Omega<20\,\text{THz}$) the Raman process can be stimulated by thermal phonons. However, at room temperature and for larger phonon frequencies, the phonon population is small, and spontaneous Raman scattering dominates. In this regime, the idler photon from SFWM is typically contaminated with light at the Stokes frequency, $\omega_S$. The Raman gain profile for silica allows significant stimulated Stokes photon production approximately 100\,nm detuned from the pump wavelength when pumping in the visible to near infrared \cite{agrawal06}. One approach to eliminate stimulated Raman contamination from the idler wavelengths is to cool the fiber \cite{takesue05, lee06, dyer08, dyer09, hall09b}. However, this is difficult in practice and adds significant experimental complications. Utilizing birefringent phase matching enables production of the idler sufficiently far from the pump to avoid contamination by Raman scatter.

\begin{figure}[h]
\centering\includegraphics[width=0.5\textwidth]{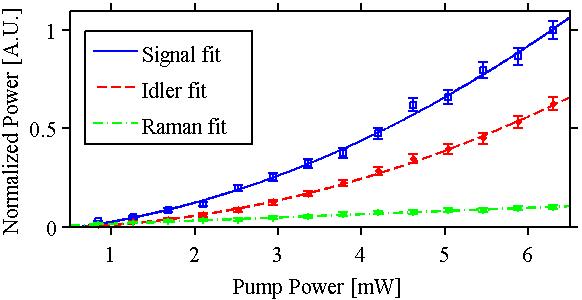}
\caption{Raman scatter (green +), signal (blue $\Box$ ) and idler (red $\diamond$) normalized power as a function of the pump power for 1\,m fiber (Fibercore HB800G) and corresponding fit lines. In the spontaneous regime shown here, the Raman scatter scales linearly with the pump power, while the SFWM scales quadratically.}
\label{fig:ramanscatter}
\end{figure}

The number of Raman photons produced in the spontaneous regime scales linearly with the pump power $P_p$, and fiber length $L$, i.e. $N_{R} \propto P_p L$. The number of idler photons scales quadratically as a function of the pump power and linearly with fiber length, i.e. $N_{i} \propto {P_p}^2 L$ \cite{lin06, lin07}. This is illustrated in Fig. \ref{fig:ramanscatter}, where the experimentally obtained signal, idler, and Raman powers as a function of the pump power are shown for a 1\,m fiber length. The signal-to-noise ratio, defined by the ratio of the number of idler photons and Raman scattered photons, is proportional to the pump power, ${\rm{SNR}} = N_i / N_R \propto P_p$. For higher pump powers the probability that detection of an idler photon originates from SFWM over Raman scattering is increased. However, for the purpose of heralding a single photon, coincidence detection events that do not originate from SFWM are the most detrimental. The most significant contribution to such coincidences arises from the detection of a signal photon and Stokes photon simultaneously. The rate of these unwanted coincidences depends solely on the probability to detect a Stokes photon per SFWM event, which scales linearly with pump power. Therefore, the Raman contamination of coincidence events actually increases with increased power. Thus, a compromise must be struck between source brightness and background noise suppression. It is therefore crucial to reduce Raman contamination by other means, such as generating photons spectrally far from the pump, which cross-polarized birefringent phase matching can achieve. 

\section{Spontaneous four-wave mixing in birefringent fibers: Experiment}
\begin{figure}[h]
\centering\includegraphics[width=0.5\textwidth]{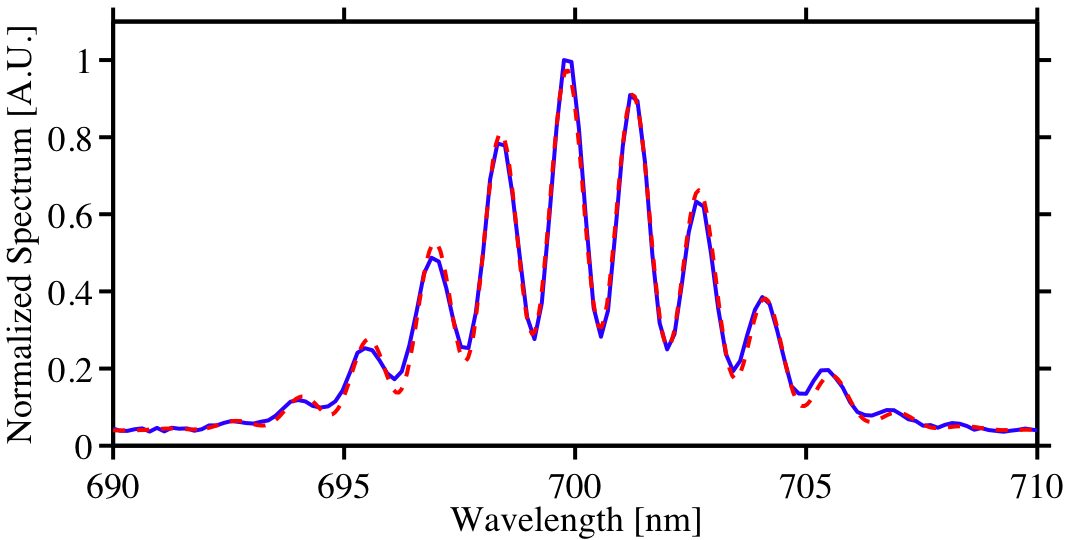}
\caption{Typical spectral fringe pattern (solid blue) and fit (dashed red) used to determine the fiber birefringence. Here the laser was tuned to 700\,nm for 0.9\,m of fiber (Fibercore HB750).}
\label{fig:birefringence}
\end{figure}

Since birefringence plays a central role in the SFWM experiment, this was measured for several commercially-available fibers over a spectral region from 690\,nm to 850\,nm. This was done by launching into the fiber broadband (10\,nm to 20\,nm) femtosecond laser pulses from a Ti:Sapphire laser (central wavelength centered at 10\,nm increments from 690\,nm to 850\,nm) plane-polarized at $45^{\circ}$ with respect to the fiber fast axis. The output light from the fiber passed through a polarizing beam splitter prior to entering a spectrometer, where spectral interference between the two polarization modes was observed, as shown in Fig. \ref{fig:birefringence}. For a fiber of length $L$ and pump with central wavelength $\lambda_0$, the birefringence can be calculated from the spectral fringe spacing, $\delta \lambda$, using $\Delta n = \lambda_{0}^{2} / (L \delta \lambda)$. The birefringence measurements showed that $\Delta n$ is constant within experimental errors over the wavelength range examined. There was a small, but not insignificant spread in the measured birefringence of the different fibers, from $3.9\times 10^{-4} \pm 0.2 \times 10^{-4}$ to $4.3\times 10^{-4} \pm 0.2 \times 10^{-4}$. Because the detuning of the signal and idler photons from the pump central wavelength is determined by the birefringence, the fiber with the largest measured birefringence (Fibercore HB800G) was used in the SFWM experiments. This ensures that photon pairs are produced as far from the pump as possible to avoid unwanted Raman contamination. The birefringence measurements are summarized in Table \ref{table}. Note that the quoted birefringence is only a lower bound given by the manufacturer.

\begin{table}[ht]
\caption{Birefringence ($\Delta n)$ Quoted (at 633\,nm) and Measured (at 690\,nm).}
\centering
\begin{tabular} {c c c c}
\hline \hline 
Manufacturer 	& Fiber Model 	& Quoted				 	& Measured \\
			&			& $\Delta n$ [10$^{-4}$]	& $\Delta n$ [10$^{-4}$] \\ [0.5 ex]
\hline
Fibercore 		& 	HB750 		& 	3.0 	& 	3.9 $\pm$ 0.2 \\
Fibercore 		& 	HB780PM 	& 	3.5 	& 	4.3 $\pm$ 0.2 \\
Fibercore 		& 	HB800G 		& 	4.2 	& 	4.3 $\pm$ 0.2 \\
3M 			& 	FS-LS-4616 	& 	4.0 	& 	4.0 $\pm$ 0.2 \\ 
\hline
\end{tabular}
\label{table}
\end{table}

\begin{figure}[h]
\centering\includegraphics[width=0.85\textwidth]{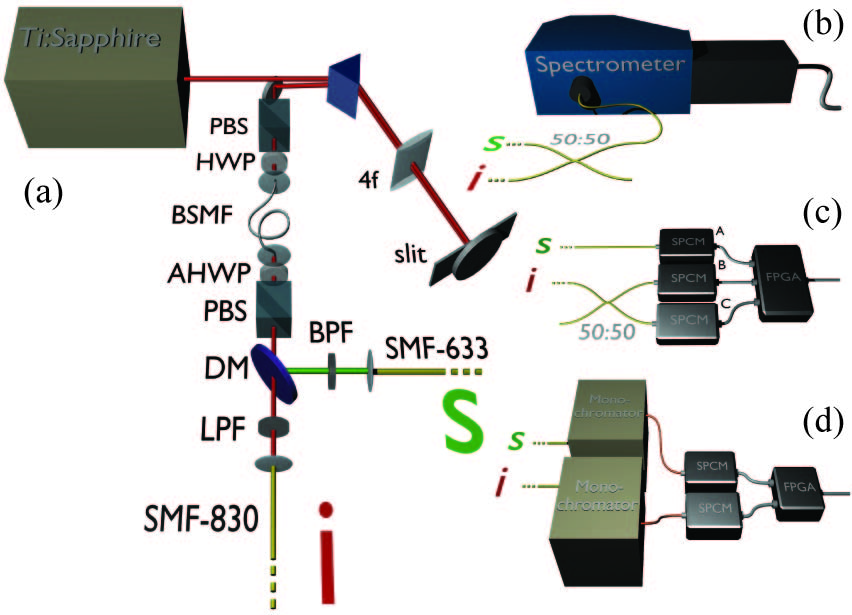}
\caption{Experimental setup. (a) The source consists of a Ti:Sapphire laser beam that is spectrally filtered using a folded 4$f$ prism pulse shaper and passes through a polarizing beam splitter (PBS) and half-wave plate (HWP) to control the polarization launched into the birefringent single-mode fiber (BSMF). An achromatic half-wave plate (AHWP) and PBS separate the pump from the SFWM, which are subsequently split at a dichroic mirror (DM) and further filtered using a band-pass filter (BPF) (long-pass filter (LPF)) in the signal (idler) beam. The signal (green, top) and idler (red, bottom) are directly coupled into single-mode fibers (SMF-633 and SMF-830), which can be directed to (b) a spectrometer via 50:50 fiber coupler to measure the marginal spectra of the photons, (c) single-photon counting modules (SPCMs) directly and via 50:50 fiber coupler to measure coincidence rates and conditional $g^{(2)}(0)$, and (d) monochromators set to $\lambda_s$ and $\lambda_i$ with outputs coupled to SPCMs via multi-mode fibers. Count rates and coincidences are registered using a field-programmable gate array (FPGA) connected to a personal computer (not shown).}
\label{fig:exp_setup}
\end{figure}

In order to demonstrate the capability of cross-polarized phase matching in standard birefringent single-mode optical fibers the experimental setup depicted in Fig. \ref{fig:exp_setup} (a) was implemented. Femtosecond pulses from a Ti:Sapphire laser (80 MHz repetition rate, 704\,nm central wavelength, and 10\,nm bandwidth FWHM) were spectrally filtered using a folded 4$f$ prism pulse shaper with a hard-edge filter (adjustable slit), resulting in a nearly square pump spectrum of 2.5\,nm bandwidth FWHM. To control the polarization of the light launched into the fiber, the shaped pulses were passed through a polarizing beam splitter (PBS) and half-wave plate before being focused into the birefringent single-mode fiber (BSMF) (10\,cm length) with an aspheric lens. The light emitted from the fiber consisting of the pump, Raman, signal and idler polarized along the slow, fast and slow, and fast axes respectively, was collimated with a second aspheric lens. To split the signal and idler from the pump an achromatic half-waveplate (AHWP) oriented the polarizations of the pump and some of the Raman vertically and the signal and idler horizontally, which were then split at a PBS. The signal and idler were split at a dichroic mirror (DM). To further suppress the pump and any Raman scatter near the pump wavelength the signal photon was passed through a broadband interference filter (IF - 36\,nm bandwidth FWHM centered at 607\,nm wavelength) while the idler was sent through an angle-tuned long-pass filter (LPF - 830\,nm transition-edge wavelength). The signal and idler photons were then coupled into single-mode fibers (SMFs) designed for 633\,nm and 830\,nm, respectively.

\begin{figure}[h]
\centering\includegraphics[width=0.85\textwidth]{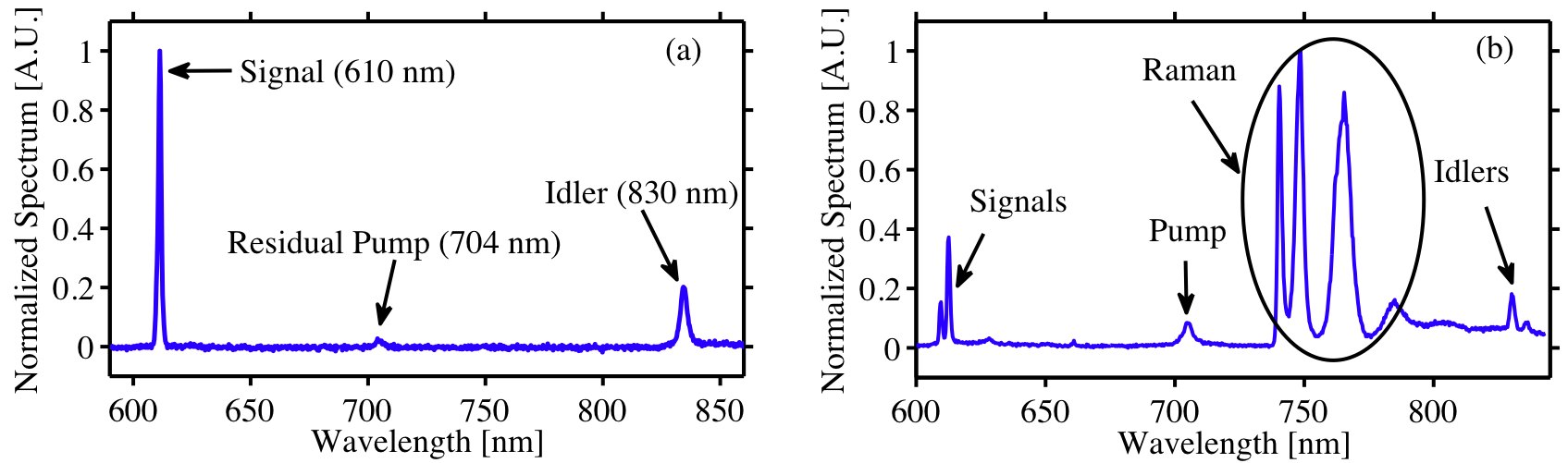}
\caption{Normalized marginal spectra obtained using the Fibercore HB800G fiber and 15\,mW average pump power. (a) Signal and idler peaks with residual pump for 3\,nm bandwidth FWHM pump centered at 704\,nm wavelength with 20\,cm fiber length. This technique was used to map out the phase matching curves as a function of the pump central wavelength as shown in Fig. \ref{fig:bi-phasematch} (b). (b) Multi-mode SFWM signal and idler peaks with residual pump and Raman scatter through an edge filter for 5\,nm bandwidth FWHM pump centered at 704\,nm with 4\,cm fiber length.}
\label{fig:marginals}
\end{figure}

Observation of the cross-polarized birefringent phase-matched SFWM was initially performed by mixing the signal and idler photons at a 50:50 fiber beam splitter and connecting one output port to a fiber-coupled spectrometer as depicted in Fig. \ref{fig:exp_setup} (b). Figure \ref{fig:marginals} (a) shows a typical spectrum obtained from such a measurement for $\lambda_{p0} = 704$\,nm, $\Delta \lambda_p \approx 3$\,nm, and 15\,mW average pump power. Note that the signal near 611\,nm is essentially free of Raman background with a signal-to-noise ratio of $\gtrsim 100$, limited by detector noise. The idler, at 830\,nm, is slightly contaminated with Raman background (signal-to-noise ratio of approximately 20). Raman background adds little to false coincidence counts, but affects the signal heralding efficiency, decreasing it from $\eta_s \approx$ 27.9\% $\pm$ 0.2\% to 26.5\% $\pm$ 0.2\%, including the detector efficiency. These relatively high efficiencies are a direct result of the standard single-mode fiber used, which allows good spatial mode matching between source and fiber networks. Considerable gains in heralding efficiency should be possible by performing all spectral filtering and sorting in fiber \cite {mcmillan09}, which is difficult with PCF. By tuning the pump central wavelength the phase-matching contours of Fig. \ref{fig:bi-phasematch} were mapped out using this experimental configuration. In some situations multimode SFWM was exhibited -- more than one SFWM peak was observed in the signal and idler spectrum as shown in Fig. \ref{fig:marginals} (b). This behavior is due to the fact that two different spatial modes were excited by the SFWM. This was attributed to poor coupling of the pump and short fiber lengths that allow non-propagating modes to be transmitted. This effect may prove useful for space-time coupling and generation of different spatial-spectral correlations via FWM in fibers, but has not been further explored.


The production rate of photon pairs, conditional second-order correlation function $g^{(2)}(0)$, and relative heralding efficiency of the SFWM photons were measured using the experimental setup depicted in Fig. \ref{fig:exp_setup} (c). The signal SMF was coupled directly to a single-photon counting module (SPCM-A), while the idler SMF was sent to a 50:50 fiber beam splitter whose outputs were coupled to a pair of SPCMs (SPCM-B and SPCM-C). The individual photodetection- and coincidence-count rates between the detectors were recorded using field-programmable gate array (FPGA) electronics connected to a personal computer. For an average input pump power of 15\,mW (704\,nm central wavelength and 3\,nm bandwidth FWHM) and fiber length of 10\,cm, individual rates of $R_s = $ 163000\,/s and $R_i = $ 87000\,/s and coincidence count rate of $R_{CC} = $ 23000\,/s were measured. This leads to raw heralding efficiencies $\eta_s = 0.265 \pm 0.002$ and $\eta_i = 0.141 \pm 0.001$, where uncertainties arise from Poisson count statistics. This is much brighter ($\approx 1600$\,pairs/s/mW) than bulk sources ($\approx 330$\,pairs/s/mW) with the same average pump power \cite{mosley08, mosley08b}, and comparable to PCF sources ($\approx 7000$\,pairs/s/mW) \cite{goldschmidt08}.

The conditional second-order correlation function $g^{(2)}(0)$ is a useful measure of the non-classical nature of a light source \cite{goldschmidt08, thorn04, uren05b, grangier86}. A value of $g^{(2)}(0)<<1$ indicates that when the signal detector fires (heralding the existence of an idler photon) the likelihood there is more than one photon in the idler mode is negligible. For classical light $g^{(2)}(0) \geq 1$, so that the degree of non-classicality of a light source is given to some extent by violation of this inequality. The various detection events obtained in 300\,s using the experimental setup shown in Fig. \ref{fig:exp_setup} (c), are given in Table \ref{table2}.
\begin{table}[ht]
\caption{Counts in 300s for $g^{(2)}(0)$ measurement}
\centering
\begin{tabular} {c c}
\hline \hline 
$N_A$ = 24652543 	& 	$N_{AB} = 1182529$\\
$N_B$ = 12353888 		& 	$N_{AC} = 1057066$ \\
$N_C$ = 11022920 		&$N_{BC} = 5940 $\\
					&$N_{ABC} = 1115$ \\ 
\hline
\end{tabular}
\label{table2}
\end{table}
\noindent The conditional second-order coherence function is given by \cite{thorn04, uren05b, grangier86}
\begin{equation}
g^{(2)}(0) =\frac{ N_{ABC}N_{A} }{ N_{AB} N_{AC} },
\label{eq:g2}
\end{equation}
\noindent where $N_{i}, N_{ij}, N_{ijk}$ are the singles, two-fold and three-fold coincidence counts in a given time for detectors $i, j, k \in \{A, B, C\}$. Using Eq. (\ref{eq:g2}) and the data in Table \ref{table2}, the second-order coherence function is calculated to be $g^{(2)}(0) = 0.022 \pm 0.001$, where errors assume Poisson count statistics. This result shows significant suppression below the classical limit of 1, indicating that the light source indeed emits photon pairs with little contamination from higher photon numbers and Raman scatter.

\begin{figure}[h]
\centering\includegraphics[width=0.85\textwidth]{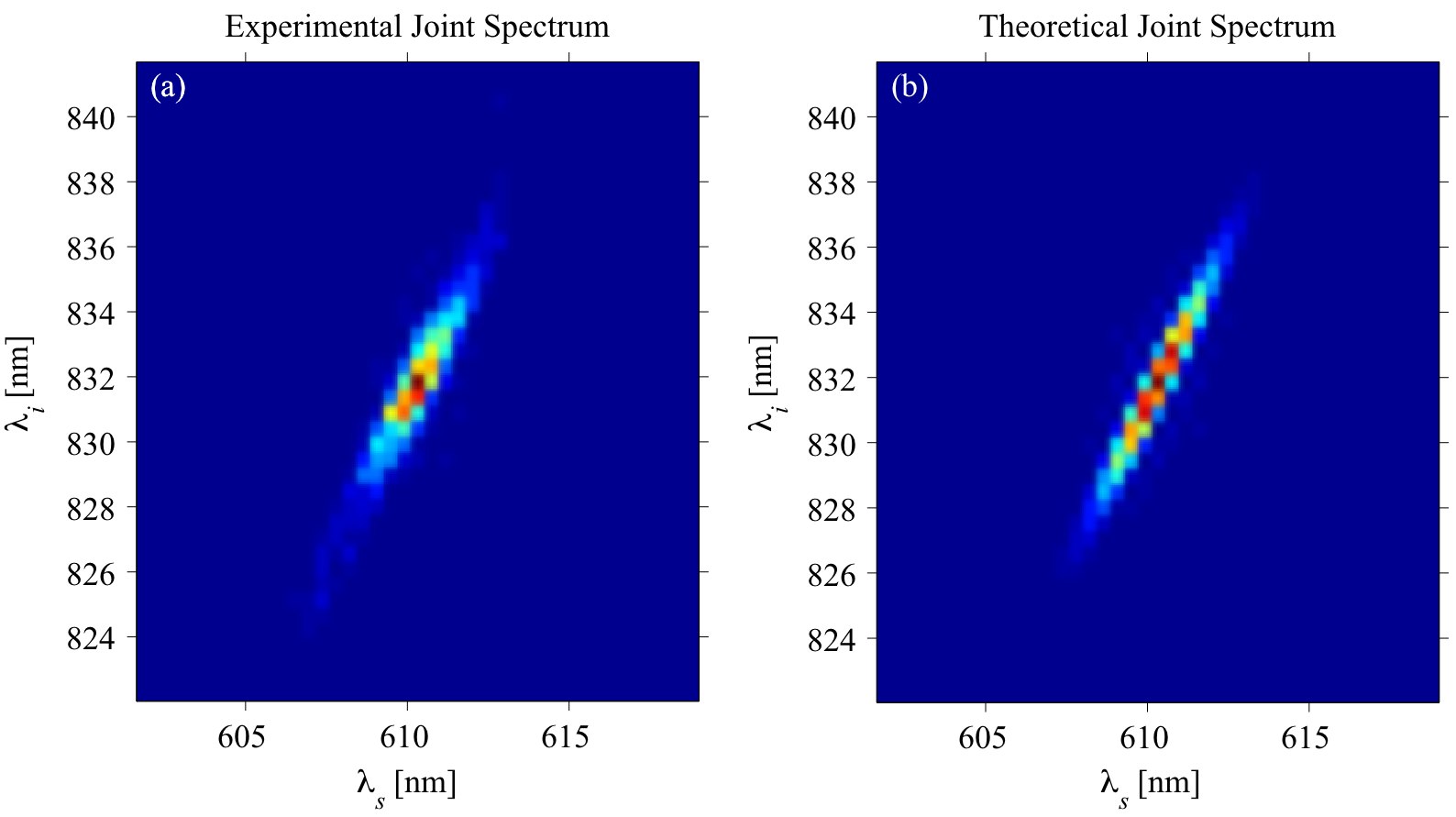}
\caption{(a) Experimental and (b) theoretical joint spectrum for a 10\,cm long fiber, 5.4\,nm bandwidth FWHM and 704\,nm central wavelength pump.}
\label{fig:jointspec}
\end{figure}

The two-photon joint spectrum was measured using the experimental setup shown in Fig. \ref{fig:exp_setup} (d). The SMFs containing the signal and idler beams were directed into two separate monochromators with 600 lines/mm gratings. The output of the monochromators were directly coupled into multimode fibers (MMFs) with 100 $\mu$m diameter, and sent to SPCMs \cite{kim05, mosley08b}. The position of the MMFs determined the central wavelength of the coupled light. The coincidence count rate between the two MMF outputs was recorded as a function of the fiber positions at the output of the monochromators, and thus wavelength settings ($\lambda_s$ and $\lambda_i$). The experimental joint spectrum for an input pump power of 14\,mW (704\,nm central wavelength and 5.4\,nm bandwidth FWHM) and fiber length $L = 10$\,cm (Fibercore HB800G) is shown in Fig. \ref{fig:jointspec} (a). The theoretical prediction for this case is shown in Fig. \ref{fig:jointspec} (b), calculated using our simplified model with birefringence, $\Delta n = 4.3 \times 10^{-4}$.

The measured and theoretical joint spectra are well-matched qualitatively. A more quantitative measure of this is the fidelity or overlap, which for two density matrices $\rho$ and $\sigma$ is given by $F(\rho,\sigma) = {\rm{Tr}}(\sqrt{\sqrt{\rho}\sigma\sqrt{\rho}})$. Assuming flat phase, this can be written in terms of the experimental and theoretical joint spectral probabilities, $p_{exp}(\omega_s, \omega_i)$ and $p_{th}(\omega_s, \omega_i)$, as \cite{nielson00}
\begin{equation}
F(p_{exp}, p_{th}) = \int \int d\omega_s d\omega_i \sqrt{p_{exp}(\omega_s, \omega_i) p_{th}(\omega_s, \omega_i)}.
\label{classical_fidelity}
\end{equation}
\noindent Using Eq. (\ref{classical_fidelity}) the overlap between the measured joint spectrum and theoretical joint spectrum is $F = 0.92 \pm 0.07$, calculated directly with no fit parameters. This implies that our model predicts the joint spectrum of the photon pair considerably well. The measured joint spectrum clearly shows frequency correlations between the signal and idler photons, indicating the joint spectral amplitude is not factorable. To quantify the degree of factorability (or de-correlation), the Schmidt number can be calculated by taking a singular-value decomposition of the joint spectrum assuming a constant phase \cite{mosley08b}. This yields a lower bound on the purity of a heralded photon, which for this two-photon spectral state is 0.22. Note that for a pump bandwidth of approximately 0.4\,nm FWHM, as calculated from Eq. (\ref{eq:pumpbandwidth}), using the fiber length and dispersion from this experiment, an approximately factorable two-photon state could be produced with this configuration. This would enable heralded pure-state photons with no need for tight spectral filtering from the fiber.


\section{Conclusion}

We have experimentally demonstrated photon-pair generation in standard birefringent single-mode fiber utilizing cross-polarized birefringent phase matching. This is a useful alternative to previous methods and offers unique advantages by producing photons in a region where waveguide dispersion can be neglected for phase matching. A model using only the material dispersion and a constant birefringence works well in the spectral region of experiments presented here. The predicted phase-matched wavelengths and joint spectral amplitude of the two-photon state produced by SFWM are well matched to the experimental measurements. The model predicts a broad range of two-photon spectral states -- ranging from correlated to completely factorable. All of this can be accomplished at wavelengths well within the spectral sensitivity of photon-counting detectors and the spectral range of a Ti:Sapphire pump laser. Indeed, a key advantage of this method is that the joint spectral shape of the produced two-photon state is relatively independent of the pump central wavelength and depends mainly on the fiber length and pump bandwidth, two relatively easy experimental parameters to control. This technique could be easily used to create heralded single-photons with narrow bandwidths and at telecommunications wavelengths by using narrower pump bandwidth and longer wavelength pump, respectively. The experimentally observed high heralding efficiencies demonstrate the potential use in integrated optical networks. This source is anticipated to be useful for a broad audience of experimentalists in quantum optics and information.

\section*{Acknowledgments}
This work was supported by the EPSRC through the QIP IRC, the EC under QAP, the IST directorate and the Royal Society. Fibercore provided fiber samples for part of this research. We thank H. J. McGuinness, M. G. Raymer, C. J. McKinstrie, and E. Brainis for useful discussions.

\end{document}